\documentclass[reprint,superscriptaddress,aps,prl]{revtex4-2}

\usepackage{graphicx}
\usepackage{dcolumn}
\usepackage{bm}
\usepackage{amsmath}
\usepackage{amsfonts} 
\usepackage{enumitem}
\usepackage{booktabs}
\usepackage{amssymb}
\usepackage{hyperref}
\usepackage{xcolor}
\usepackage{siunitx}
\begin{document}

\title{Self-organized adaptive branching in frangible matter}    

\author{P.L.B. Fischer}
\thanks{These authors contributed equally to this work.}
\affiliation{School of Engineering and Applied Sciences, Harvard University, Cambridge, MA 02138, USA}

\author{J. Tauber}
\thanks{These authors contributed equally to this work.}
\affiliation{School of Engineering and Applied Sciences, Harvard University, Cambridge, MA 02138, USA}

\author{T. Koch}
\affiliation{Department of Scientific Computing and Numerical Analysis, Simula Research Laboratory, Oslo, Norway}

\author{L. Mahadevan}
\email{lmahadev@g.harvard.edu}
\affiliation{School of Engineering and Applied Sciences, Harvard University, Cambridge, MA 02138, USA}
\affiliation{Department of Organismic and Evolutionary Biology, Harvard University, Cambridge, MA 02138, USA}
\affiliation{Department of Physics, Harvard University, Cambridge, MA 02138, USA}

\begin{abstract}
Soft and frangible materials that remodel under flow can give rise to branched patterns shaped by material properties, boundary conditions, and the time scales of forcing. We present a general theoretical framework for emergent branching in these frangible (or threshold) materials that switch abruptly from resisting flow to permitting flow once local stresses exceed a threshold, relevant for examples as varied as dielectric breakdown of insulators and the erosion of soft materials. Simulations in 2D and 3D show that branching is adaptive and tunable via boundary conditions and domain geometry, offering a foundation for self-organized engineering of functional transport architectures.
\end{abstract}

\maketitle

Flow-driven self-organization of branched structures is widespread in nature, encompassing phenomena such as dielectric breakdown in insulating materials~\cite{Blanchini2021}, river delta formation~\cite{Konkol2022}, slime mold growth~\cite{Ghanbari2023}, channel formation in ant colonies~\cite{Ocko2015}, and erosion of frangible porous media~\cite{Mahadevan2012,Derr2020}. A unifying feature of many such systems is an external field that drives flow (of fluid, current, heat, etc.) through a material that transitions sharply from resisting to permitting flow once local stresses exceed a threshold. This leads to similar branching patterns in what are otherwise very different physical systems, raising the question: is there a unified theoretical approach to these problems? Here we propose a general framework to model flow-driven branching in such frangible or \emph{threshold materials} by coupling a conservation law with a minimal description for the local material response. We also point out the critical role of boundary conditions in controlling the morphology of these patterns and enabling their functional control in engineered and living systems.


\begin{figure}[!htb]

    \centering
    \includegraphics[width=\linewidth]{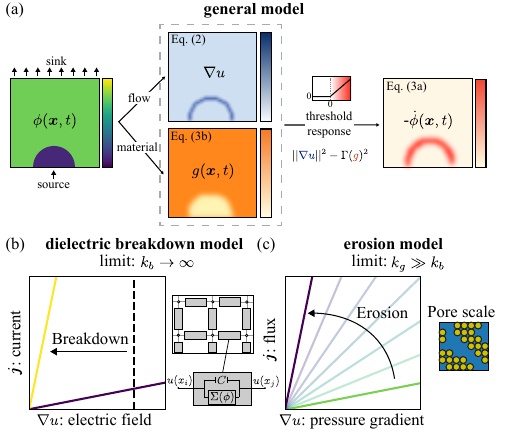}
    
    \caption{Dynamics of frangible/threshold materials: (a) The local material state $\phi$ evolves under the combined influence of flow and material properties. Flow affects the material through the gradient field $\nabla u$ (Eq.~\ref{eq:flowmass}), while non-local material interactions are described by the field $g$ (Eq.~\ref{eq:nonlocal}). Together, these fields drive the threshold-like evolution of $\phi$ (Eq.~\ref{eq:ReLU}), leading to the expansion of high-conductivity zones and the emergence of branched patterns in the presence of disorder. (b) Dielectric breakdown represents a fast, binary state limit of the general model. In our implementation on a network of capacitors and non-linear resistors (Eq.~\ref{eq:dielectric_closure}), edges switch from insulating (yellow) to permanently conductive (purple) once the voltage threshold $\Gamma$ is exceeded. (c) Flow-driven erosion represents a slow, gradual limit, where the solid fraction $\phi$ of a porous medium is progressively degraded, enhancing local conductivity (Eq.~\ref{eq:erosion_closure}). 
    }
    
    \label{fig:models}
    
\end{figure}

We model the mesoscopic state of the material using a dimensionless field $\phi(\boldsymbol{x},t) \in (0,1)$ evolving in space and time; this field (or its complement) indicates regions where the material has locally changed properties and become a conductor (of matter, energy or information). Flow through the material is governed by a simple relationship between the flux $\boldsymbol{j}$, a macroscopic potential $u$, and its rate of change $\dot{u} := \partial_t u$ as given by
\begin{align}
    \boldsymbol{j} &= -\Sigma(\phi) \nabla u - C \nabla \dot{u} \;. \label{eq:flow}
\end{align}
The first term describes linear flow (\textit{e.g.}, Darcy's law or Ohm's law) with conductivity $\Sigma(\phi)>0$ depending on the local material state. The second term captures time-dependent energy storage and release in the surrounding medium via a capacitance $C\geq 0$ or an effective (visco-)elasticity. Local conservation (\textit{e.g.}, mass conservation or charge conservation) requires $\nabla \cdot \boldsymbol{j} = 0$, leading to
\begin{align}
    \nabla \cdot \left(\Sigma(\phi) \nabla u + C \nabla \dot{u}\right)&= 0 \;. \label{eq:flowmass}
\end{align}

To close the system, we need to describe how the material response $\phi({\bf x},t)$ evolves in response to local fields, \textit{e.g.}, mechanical or electric stresses. Inspired by the rapidity of processes associated with dielectric breakdown, failure of fuses, and erosive processes, we define a threshold material as one for which the local evolution equation for $\phi(\boldsymbol{x},t)$ can be written as:
\begin{subequations}
\begin{align}
   \dot{\phi}(\boldsymbol{x},t) &=  -k_b\phi \,\Gamma_0^{-2}\operatorname{ReLU}\lbrace ||\nabla u||^2 - \Gamma(g)^2 \rbrace \;, \label{eq:ReLU}\\
    \dot{g}(\boldsymbol{x},t) &= \nabla \cdot (D\nabla g) + k_g(\phi - g) \;\;, \label{eq:nonlocal}
\end{align}
    \label{eq:General}
\end{subequations}
where $k_b$ sets the degradation rate, $\Gamma_0$ is a characteristic gradient scale, and $\Gamma(g)$ is a material-specific threshold function modulated by the non-local field $g(\boldsymbol{x},t)$. This general formulation can be implemented either on a discrete network or in a continuum setting.
The $\operatorname{ReLU}$ (Rectified-Linear Unit) function enforces a threshold response, allowing evolution only when the local gradient magnitude $\|\nabla u \|$ exceeds $\Gamma(g)$. We note that 
this form is very similar to the activation functions underlying neural networks, but here it is specified in terms of spatiotemporally varying fields rather than an aggregation of discrete inputs. The quadratic form ensures symmetry under sign changes in $\nabla u$, based on elementary physical considerations, as introduced in~\cite{Mahadevan2012}. The field $g$, inspired by granular media transport models~\cite{Kamrin2024}, is an auxiliary field coupled to the evolving material state $\phi$ through diffusion ($D$) and relaxation ($k_g$). It captures how local structure is influenced by short-range interactions and the recent spatiotemporal history of $\phi$. This non-local representation enables the gradient sensitivity threshold $\Gamma(g)$ to reflect local variations; when the degradation rate $k_g$ is very large (appropriately scaled), we recover previous models~\cite{Mahadevan2012,Derr2020}.
The dependence of both $\Sigma$ and $\dot{\phi}$ on $\phi$ creates a self-reinforcing feedback loop that drives the flow-induced expansion of high conductivity zones~\cite{Derr2020}. This process is highly sensitive to disorder, leading to branched patterns triggered by spatial heterogeneity in the material field $\phi$ or the threshold function $\Gamma$. 

Physical systems correspond to different limits of this general model, which is governed by three key timescales: the flow relaxation rate $k_r \sim \Sigma_0 / C$, with $\Sigma_0$ the conductivity for $\phi(\bm{x},0)$, the material degradation rate $k_b$, and the material short-range interaction rate $k_g$. Their relative magnitudes determine whether degradation is instantaneous, local, or mediated by non-local structural dynamics. In the following, we explore two threshold systems, dielectric breakdown and erosion, that reflect distinct asymptotic regimes of this model, yet---as we demonstrate below---exhibit analogous branching behavior (see Fig.~\ref{fig:models} and Table S1~\cite{supp}).

Dielectric breakdown, observed in phenomena like lightning and fractal woodburning, produces arborized Lichtenberg figures within seconds~\cite{Takahashi1979}. Breakdown occurs when the local electric field strength surpasses a threshold, rendering the medium locally conductive and allowing current to flow along the newly formed branches~\cite{Hager1989}. Since breakdown occurs much faster than the flow field can adjust, this corresponds to the asymptotic limit $k_{\text{b}} \to \infty \gg k_{\text{r}}$, in which material degradation is instantaneous. 
To model this limit, we use a minimal network-based framework with a binary material state $\phi(\boldsymbol{x},t)\in \{0,1 \}$ encoding breakdown, inspired by recent work on lightning strikes~\cite{Blanchini2021} (Fig.~\ref{fig:models}(b)). The dielectric breakdown model is detailed in the End Matter, Appendix~A. 

Erosion in porous media such as sand beds is an example of slow self-organized branching in a frangible or threshold material, occurring over timescales ranging from minutes to years~\cite{Mahadevan2012,Derr2020}. When flow-induced stress exceeds the breaking stress of grains, the porous structure reconfigures and its permeability changes. In this regime, flow redistributes much faster than erosion progresses, so that $ k_{\text{r}} \gg k_{\text{b}}$. 
To model this regime in a continuum description, we take the limit  $C\rightarrow 0$, reducing Eq.~\eqref{eq:flow} to $\nabla \cdot \left(\Sigma(\phi) \nabla u\right)= 0$, with $u(\boldsymbol{x},t)$ the fluid pressure and $\phi(\boldsymbol{x},t)\in (0,1)$ the solid volume fraction averaged over the pore scale (Fig.~\ref{fig:models}(c)). The erosion model is detailed in the End Matter, Appendix~B. 

\begin{figure}

    \centering
    \includegraphics[width=\linewidth]{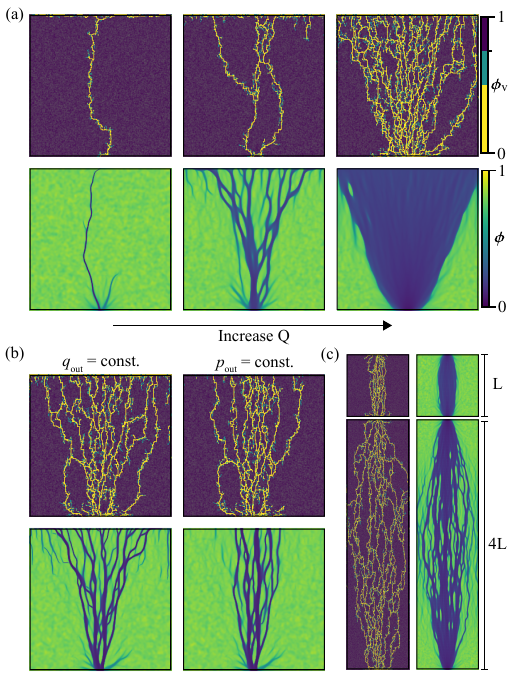}
    
    \caption{Boundary conditions affect branch shape: We visualize the steady-state $\phi$-field from numerical solutions of the dielectric breakdown model (End Matter, Appendix~A) and erosion model (End Matter, Appendix~B). For dielectric breakdown, $\phi_v$ is the vertex-averaged $\phi_e$ of connected edges. (a) Increasing the flux amplitude raises both the number of branches and their width. Parameters: dielectric breakdown $Q \in \{0.01,0.1,1.0\},\ T = 1.0$; erosion $Q \in \{0.05,0.5,5.0\},\ T=10, \ \xi = 0.05$. (b) Comparison between homogeneous flux sink (left) and homogeneous potential or pressure sink (right). Parameters: $Q = 0.5, \ T=1.0$ for dielectric system, $Q = 0.5, \ T = 10, \ \xi =0.025$ for eroding system. (c) Increasing the source-sink distance by a factor of 4 leads to more hierarchical branches that span a larger region. Parameters: dielectric $Q = 0.5, \ T = 1.0$, erosion $Q = 2, \ T = 10, \ \xi = 0.05$.} 
    \label{fig:boundarycontrol}
    
\end{figure}
In both models, an externally imposed flow from a source will drive the self-organization of branched patterns, eventually connecting to a sink. Notably, in threshold materials, the irreversible nature of the conductivity change allows the branches to persist once formed. The shape of the mature structure is determined by the interplay of the threshold material's local structural heterogeneity and the branches' out-of-equilibrium evolution. The out-of-equilibrium evolution is controlled by the rate-limiting material time scales, $k_r$, $k_b$, and $k_g$, and the boundary conditions.

To characterize the range of possible patterns, we begin with a configuration featuring a point source and a line sink on opposing sides, with zero flux conditions on all other boundary regions. At the source, we apply a time-dependent flux $q_{in}(t)$
\begin{align}
    q_{in}(t) &= Q \min\left\lbrace 1, t/T \right\rbrace,
    \label{eq:flowramp}
\end  {align}
where $Q$ and $T$ are variable parameters. As shown in Fig.~\ref{fig:boundarycontrol}(a) (see also Video S1), increasing $Q$ promotes branching in both models. This behavior can be attributed to the increase in the ratio $\tau / T_f$, where $\tau$ is the material response time and $T_{\text{f}} \sim Lw_c/Q$ the flow timescale over which fluid fills a channel of width $w_c$ and length $L$. Similarly, branching is enhanced when $\tau / T$ is increased (see  Supplementary Material~\cite{supp}). Next, we switch the boundary condition at the sink from homogeneous flux to homogeneous pressure, with results as shown in Fig.~\ref{fig:boundarycontrol}(b). This change leads to less widespread branching patterns than the flux condition, as homogeneous pressure does not require an equal flux to reach every point on the boundary. 
Finally, we explore the impact of source-sink separation, as shown in Fig.~\ref{fig:boundarycontrol}(c). As the separation increases, more pronounced arborized structures emerge. This transition is governed by the ratio $\xi/L$ of the communication length $\xi\sim\sqrt{D/k_g}$ and the size of the system $L$ as can be seen in Fig.~\ref{fig:boundarycontrol}(c). Here, the bottom panel shows an elongated system with a smaller ratio $\xi / 4L$, demonstrating stronger arborization compared to the system of regular size in the upper panel. Equivalently, scaling both $\xi' =  \xi/4$ and $L'= 4 L$ leaves the ratio of length scales invariant, again leading to strong arborization (see~\cite[Fig.~S2]{supp}).

Despite the differing dominant timescales, boundary conditions have a similar effect on erosion and dielectric breakdown. In both cases, the combined effect of flow, material heterogeneity, and the irreversible decay of the material structure actively pushes the system out of equilibrium, leading to the exploration of suboptimal paths and eventually the formation of permanent channels. In dielectric breakdown, the system's capacitance prevents instant equilibration, while in porous flow it is the finite time scale of erosion. The ratio between material response time and boundary-driven flow dynamics thus serves as an effective control parameter, determining how far the system is driven from equilibrium. It is interesting to note that at steady state the material response function satisfies an eikonal equation, $||\nabla u||^2 = \Gamma(\boldsymbol{\phi({\bf x}}))^2$. The solutions correspond to shortest-time paths that can be interpreted as effective transport routes through a disordered medium towards which the non-equilibrium dynamics gradually converge.

\begin{figure}

    \centering
    \includegraphics[width=\linewidth]{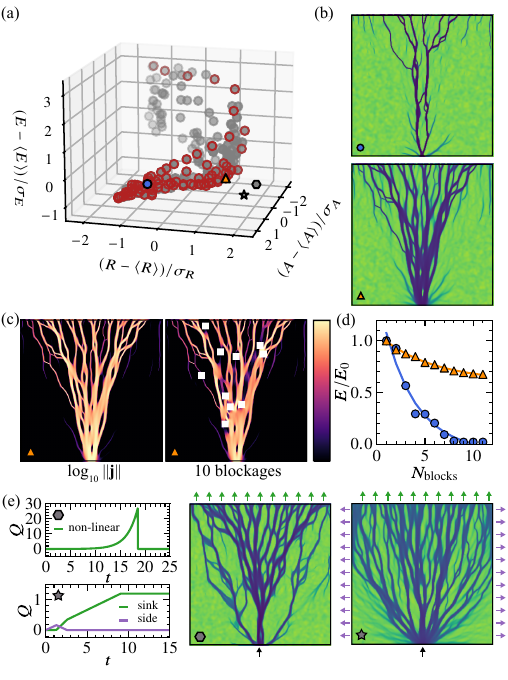}

    \caption{Function of branched structures: (a) Scaled $A$, $E$, and $R$ from steady-state $\phi$-fields of the erosion model (End Matter, Appendix~B) for $Q=[0.05,5]$, $T=[1,100]$ and $\xi=0.025$ (gray dots). Red circles mark the Pareto front. (b) Example structures with low (blue circle, $N_{\text{loops}}=12$) and high robustness (orange triangle, $N_{\text{loops}}=42$), see (a) for reference. (c) Robustness is probed by imposing random blockages ($\phi = 1$, white squares). The flux magnitude $\log_{10} \| \boldsymbol{j} \|$ shows the effect of these perturbations. (d) Normalized flux leaving the system at the sink versus the number of blockages, averaged over 5 random seeds. (e) Increased complexity of boundary control: Exponential flow-profile (Eq.~S8, $Q=0.004$, $T=2.1$, $\alpha=18.5$) yields $N_{\text{loops}}=58$ (grey star). Side boundaries as a sink (polygon) at the start of invasion (Eq.~S9, $Q=1.0$, $T=10.0$, $a=0.15$) yields $N_{\text{loops}}=56$ (polygon).
    }

    \label{fig:function}

\end{figure}

Given the emergence of self-organized arborization across a range of local threshold response functions and boundary conditions, a natural question is whether we can steer these structures toward functional patterns. To illustrate this idea, we draw inspiration from biological transport networks, which balance competing demands such as material economy (minimal construction and maintenance cost), transport efficiency, and robustness through structural trade-offs~\cite{Ronellenfitsch2019}. In our system, these costs are captured by three metrics: structure preservation ($A$), the average mass fraction $\langle \phi \rangle$ of the remaining material; efficiency ($E$), the global conductivity ($E = Q / \Delta P$); and robustness ($R$), the sensitivity to local clogging. 
We vary the boundary parameters $Q$ (maximum flux) and $T$ (time to reach maximum flux) to generate a collection of branched structures (Fig.~\ref{fig:function}(a)-(b)) and evaluate their $E$ and $A$. Although $R$ can be measured directly by introducing clogs, as shown in Fig.~\ref{fig:function}(c)-(d), the analysis is not straightforward for all structures. Since $R$ correlates with the number of loops $N_{\text{loops}}$ in the network, and loops are associated with robustness of biological transport networks~\cite{eleni}, we use $N_{\text{loops}}$ as a proxy for robustness ($R\sim N_{\text{loops}}$). 

The scaled function space shown in Fig.~\ref{fig:function} reveals that a subset of the self-organized networks lies on a Pareto front (red points), the trade-off boundary where maximizing one objective reduces another (see Supplemental Material~\cite{supp}). Along this front, $R$ increases with $E$, while $A$ gradually decreases, reflecting increasing branch complexity (Fig.~\ref{fig:function}(b)). Beyond the maximum in $R$, both $R$ and $A$ drop sharply, corresponding to washout (Fig.~S1~\cite{supp}). Increasing the complexity of boundary control can further enhance functionality: both exponential flow ramps and time-dependent modulation of out-flux across multiple sinks allow the Pareto front for linear ramps in Fig.~\ref{fig:function}(a) to be surpassed, while also promoting invasion of a larger portion of the domain (Fig.~\ref{fig:function}e and Video~S1).

To interpret the trade-off between material cost, efficiency, and robustness in threshold materials (Fig.~\ref{fig:function}(a)), it is helpful to compare them with common trends in biological transport networks~\cite{Ronellenfitsch2019,Tero2010,Bebber2007,Blinder2010}. In both cases, large-scale structures emerge from local adaptive feedback. Furthermore, analogous to our findings, adaptive flow models for biological networks show that structure is tuned by the interplay of material time scales (set by local conductivity-based feedback) and boundary time scales (set by growth or fluctuations~\cite{Ronellenfitsch2016,Ronellenfitsch2019}). In these models, steady-state structures arise from an active balance between flow-based reinforcement and restorative mechanisms. By contrast, in gradient-sensitive threshold materials steady states emerge passively, without any restoring component, once local stresses drop below a critical value. This difference is reflected in the correlation between efficiency and robustness: in biological networks, loops impose maintenance costs that shift the active balance, generating a trade-off along the Pareto front, while in threshold materials loops incur no cost once formed, breaking this interdependence. 
Overall, Fig.~\ref{fig:function} highlights that threshold materials can be guided to self-organize into functional transport networks, via dynamics fundamentally different from the active adaptation seen in biological systems. How the functionality of these networks compares quantitatively to biological ones remains an open question, as any comparison is highly context-dependent.

\begin{figure*}[!htb]

    \centering
    \includegraphics[width=\linewidth]{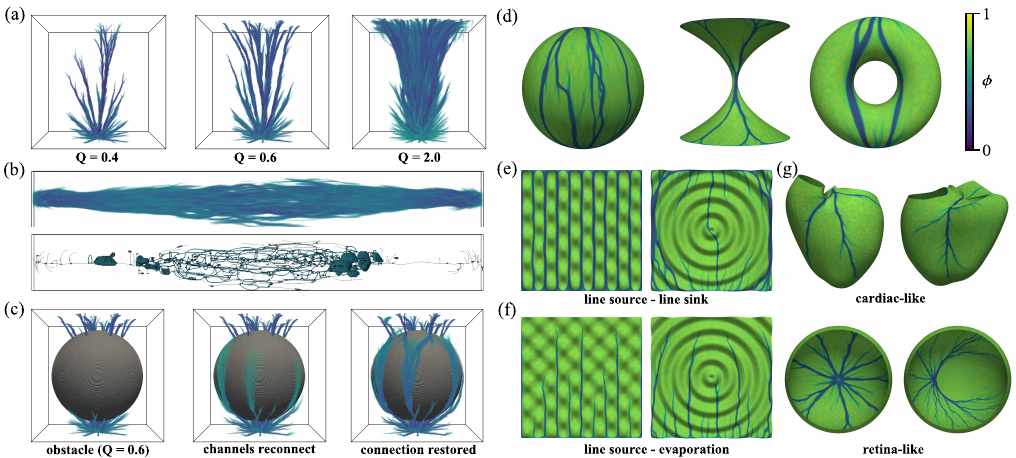}

    \caption{Guided flow-driven branching by erosion in three dimensions. $\phi$-fields from numerical solutions of \eqref{eq:flowmass} and \eqref{eq:General} with $C=0$ and model parameters given in the Supplemental Material~\cite{supp}. (a) Dependence on $Q$ for a point source-planar sink geometry. Color shows $\phi$, with transparency increasing rapidly around $\phi=0.4$. (b) Flow network formation in long domains. Top: $\phi$-field as in (a). Bottom: skeleton (see Supplemental Material~\cite{supp}), revealing loops (blue cycles) and voids (blue blobs) toward the domain center. (c) Reconnection after insertion of a blockage demonstrates resilience. (d-e) Branching on curved surfaces. (f-g) Branching on curved surfaces with areal sink terms. Cardiac surface geometry from \cite{Finsberg_fenics-beat_2024,PANKEWITZ2024103091}.}

    \label{fig:CurvedSurfaces}

\end{figure*}

The presented framework extends beyond the planar geometries explored so far. Solving the governing equations Eq.~\ref{eq:flowmass}-\ref{eq:nonlocal} for erosive threshold materials in 3D (see Supplementary Material~\cite{supp} for details), we find that branched patterns again emerge for a point source and planar sink with homogeneous flux (Eq.~\ref{eq:flowramp}), as shown in Fig.~\ref{fig:CurvedSurfaces}(a) and Video~S2. Similar to the two-dimensional case, branching in three-dimensional domains can be modulated through boundary control: increasing the source-sink separation promotes loop formation (Fig.~\ref{fig:CurvedSurfaces}(b)).
The nature of the self-organizing patterns in threshold materials naturally makes the networks resilient. When the transport network is damaged and flow is inhibited in a particular region, the dynamical system causes the pattern to reorganize the network to restore transport, as seen in Fig.~\ref{fig:CurvedSurfaces}(c) (see also Video~S2, and Supplemental Material~\cite{supp}). Thus, the system exhibits self-healing without external intervention or explicit repair mechanisms.  

We finally probe the consequences of our model when moving from Euclidean to non-Euclidean geometries, such as spheres or hyperboloids; this implies that the gradient and divergence terms in Eqs.~\ref{eq:flow}-\ref{eq:nonlocal} are replaced by surface differential operators.  We observe, not surprisingly, that domain curvature influences the spacing and orientation of branches (Fig.~\ref{fig:CurvedSurfaces}(d) and Video~S3). As suggested by our earlier discussion of the planar case, flow paths align with geodesic solutions of an eikonal equation, such that channelization tends to avoid regions of high Gaussian curvature. When we allow fluid to leave the domain through evaporation, $\nabla \cdot \left( \Sigma(\phi) \nabla u \right) = s$, $s>0$, as previously explored in flat domains~\cite{Derr2020}, a competition arises between filling space and following curvature-guided flow (Video~S4). The same principle now yields branched patterns that qualitatively resemble natural vasculature on biology-inspired surfaces, such as the retina and cardiac muscle. While we do not suggest a mechanistic link to vasculogenesis and angiogenesis, these findings indicate a clear potential use of self-organized branching as a technology to engineer perfusable network structures for biomedical purposes.

We conclude by pointing out that our minimal model for threshold materials, spanning a range of natural systems, can support the self-organization of branching flow networks, with boundary conditions enabling control over their functional architecture. A critical next step is to test our predictions in existing quasi-2D experimental systems~\cite{Mahadevan2012,Tauber2024}, or on curved surfaces and fully three-dimensional domains. Together, this outlines a framework for creating complex branched structures through self-organized engineering where structures emerge through guided dynamics rather than predefined blueprints.

\begin{acknowledgments}
We acknowledge financial support from the EMBO postdoctoral fellowship ALTF 2022-7, the Harvard NSF 20-, the Simons Foundation, and the Henri Seydoux Foundation. We thank Nick Derr for advice on implementing the erosion model, and Sumit Sinha and Vishaal Krishnan for insightful discussions on both models. We thank Henrik Finsberg for providing code for constructing the cardiac mesh.
\end{acknowledgments}


\nocite{Blanchini2021,Balay2025,Derr2020,Baratta2023,Koch2021,Finsberg_fenics-beat_2024,PANKEWITZ2024103091,Geuzaine2009,Ronellenfitsch2019}

%

\section{End Matter}
\subsection{Appendix A - Dielectric breakdown}\label{app:1}

The medium has constant capacitance $C$ and a conductivity $\Sigma$ that increases irreversibly from a small background value $\epsilon$ to a high value $r$ when the local electric field strength exceeds a threshold $\Gamma$. The binary material state $\phi(\boldsymbol{x},t)\in \{0,1 \}$ encodes this transition. We express the system in nondimensional form using $t = 1/k_r \tilde{t}$, $\nabla u = \Gamma_0 \widetilde{\nabla u}$, $\ell= \ell_0 \tilde{\ell}$ where $k_r=r/C$ and $u(\boldsymbol{x},t)$ is the electric potential. The resulting model corresponds to the instantaneous degradation limit of Eq.~\ref{eq:General}, with $k_b \to \infty$:
\begin{subequations}
\begin{align}
    \Sigma(\phi) &= \tilde{\epsilon} + (1-\phi) \;,\\
    \phi^+ &= \phi^- - \phi^- \cdot \Theta\lbrace|\widetilde{\nabla u}\|^2 - \widetilde{\Gamma}^2(\boldsymbol{x})\rbrace\;. 
    \end{align} \label{eq:dielectric_closure}
\end{subequations}
Here, $\tilde{\epsilon} = \epsilon/r \ll 1$ is the dimensionless background conductivity, $\phi^{\pm}=\phi(\boldsymbol{x},t^{\pm})$ the material state just before ($t^-$) and after ($t^+$) instantaneous breakdown, and $\widetilde{\Gamma}(\boldsymbol{x})$ a static, spatially heterogeneous threshold. When the non-dimensional electric field strength $\|\widetilde{\nabla u}\|$ exceeds this threshold, the Heaviside function $\Theta$ triggers a local, irreversible transition to a high-conductivity state ($\phi=0$) (cf.~Fig.~\ref{fig:models}(b)). A detailed explanation of the implementation for a network with a square grid is provided in the Supplementary Material~\cite{supp}. 

\subsection*{Appendix B - Erosion model}\label{app:2}

Following~\cite{Derr2020}, we nondimensionalize the equations using $t = 1/k_b \tilde{t}$, $\nabla u = (B_0/\ell_g) \widetilde{\nabla u}$, $\ell= \ell_g (B_0/(\eta k_b)) \tilde{\ell}$, where $B_0$ is a characteristic erosion stress, $\ell_g$ the grain size, $\eta$ the viscosity, and $k_b$ the erosion rate. The resulting model corresponds to a limiting case of Eq.~\eqref{eq:General}, where $k_r \to \infty$ and $k_g \gg k_b$. In this limit, the non-local field $g$ rapidly equilibrates to a spatially smoothed version of $\phi$, with interaction range $\xi=\sqrt{D/k_g}$, yielding a non-local threshold response.
\begin{subequations}
\begin{align}
    \Sigma(\phi) &= \phi^{-2}\left(1-\phi\right)^3 \;, \label{eq:CarmanKozeny}\\
    \dot{\phi} &= - \phi \:  \operatorname{ReLU}\lbrace||\widetilde{\nabla u}||^2 - \psi(g)\rbrace \;, \label{eq:ErosionThreshold}\\
    g(\boldsymbol{x}) &=\int_V \frac{\phi(\boldsymbol{y})}{2\pi \xi^2} \exp\left\lbrace-\frac{||\boldsymbol{x} - \boldsymbol{y}||^2}{2\xi^2}\right\rbrace \, \mathrm{d} \boldsymbol{y}. \label{eq:ErosionCohesion}
\end{align} \label{eq:erosion_closure}
\end{subequations}
Eq.~(\ref{eq:CarmanKozeny}) is the dimensionless Carman-Kozeny equation, linking hydraulic conductivity to solid volume fraction~\cite{Scheidegger1957}, see~Fig.~\ref{fig:models}(c). 
Eq.~(\ref{eq:ErosionThreshold}) governs erosion, with $\psi(g)$ a smoothed step function that modulates sensitivity to pressure gradients:
\begin{subequations}
\begin{align}
    \psi(g)&=(H(g)-H(0)/(H(1)-H(0)) \;\;, \\
    H(g) &= \frac{1}{2} \left(1 + \tanh\left(\omega(g - g^*)\right)\right) \;\;, 
\end{align} 
\end{subequations}
where parameters $g^*$ and $\omega$ set the threshold location and sharpness. Writing $\Gamma^2=\Gamma_0^2\psi(g)$ makes the connection to~Eq.~(\ref{eq:ReLU}) explicit. 
Eq.~(\ref{eq:ErosionCohesion}) defines $g(\boldsymbol{x})$ as a spatial convolution of $\phi(\boldsymbol{x})$ in 2D with a Gaussian kernel of communication length $\xi$, representing short-range material cohesion. 
Implementation details are provided in the Supplementary Material~\cite{supp}. 


\clearpage

\end{document}